# Enhancement of spin Hall conductivity in W-Ta alloy

Jun-Young Kim[1,2], Dong-Soo Han[1,3]*, Mehran Vafaee[1,4], Samridh Jaiswal[1,4], Kyujoon Lee[1], Gerhard Jakob[1], Mathias Kläui[1]*

[1] *Institute of Physics, Johannes Gutenberg University, Mainz 55128 Germany*
[2] *Max Planck Institute for Intelligent Systems, Stuttgart 70569 Germany*
[3] *Korea Institute of Science and Technology, Seoul 02792, Republic of Korea*
[4] *Singulus Technologies AG, 63796 Kahl am Main, Germany*

Generating pure spin currents via the spin Hall effect in heavy metals has been an active topic of research in the last decade. In order to reduce the energy required to efficiently switch neighbouring ferromagnetic layers for applications, one should not only increase the charge-to-spin conversion efficiency but also decrease the longitudinal resistivity of the heavy metal. In this work, we investigate the spin Hall conductivity in $W_{1-x}Ta_x$/CoFeB/MgO (x = 0 – 0.2) using spin torque ferromagnetic resonance measurements. Alloying W with Ta leads to a factor of two change in both the damping-like effective spin Hall angle (from - 0.15 to - 0.3) and longitudinal resistivity (60 - 120 $\mu\Omega$ cm). At 11% Ta concentration, a remarkably high spin Hall angle value of – 0.3 is achieved with a low longitudinal resistivity 100 $\mu\Omega$ cm, which could lead to a very low power consumption for this W-based alloy. This work demonstrates sputter-deposited W-Ta alloys could be a promising material for power-efficient spin current generation.

* Author to whom correspondence should be addressed: dshan@kist.re.kr & klaeui@uni-mainz.de

For the last decade, spin current generation via the spin Hall effect (SHE)[1–5] has been at the heart of spintronics research, where its application ranges from fundamental research to next generation spintronic devices. Among relevant phenomena, spin Hall effect and spin-orbit torques (SOTs) in magnetic multilayers of heavy-metal (HM)/ferromagnets (FM)[4,6–8] have attracted particular interest in the community, as they hold promise for ultra-fast and highly efficient manipulation of magnetic entities in ferromagnets[6,7,9] as well as antiferromagnets[10–12]. In particular, random access memory devices operated via spin-orbit torques (SOT-MRAM) have become a promising candidate to replace the current in-production spin-transfer-torque-MRAM providing better performances and new functionalities[13,14]. However, several challenges remain for practical use of SOT-based devices. The most challenging issue is the high current densities required for the magnetization switching by SOTs. In this context, many efforts have been put forward to reduce the energy required to switch magnetization and improve the efficiency of SOTs.

The critical current density for the switching of in-plane magnetized free layers, for example, is inversely proportional to the spin Hall angle $\theta_{\text{SH}} \equiv \left(\frac{2e}{\hbar}\right) J_{\text{SHE}}/J_{\text{c}}$, describing the conversion efficiency of the SHE-induced spin current ($J_{\text{SHE}}$) from the applied charge current ($J_{\text{c}}$). Accordingly, the threshold energy for the switching is proportional to $\propto R/\theta_{\text{SH}}^2$. Most of the research relevant to SOTs has been focused on enhancing $\theta_{\text{SH}}$ and searching for suitable materials with a large $\theta_{\text{SH}}$. Starting from conventional high $\theta_{\text{SH}}$ materials such as Pt, W and Ta, alloying[15–20] and oxygen incorporation[21] have been tried to increase $\theta_{\text{SH}}$. However, increasing $\theta_{\text{SH}}$ and spin Hall conductivity by alloying often results in an increase in longitudinal resistivity $\rho_{xx} \propto R$ at the same time, keeping the power consumption high for device operation. Hence, the current efforts are given to increase the spin Hall conductivity without a large increase of longitudinal resistivity.

β-W in the A15 crystal structure, known from many high critical-current superconductors such as Nb$_3$Sn[22], has been found to show a large spin Hall angle of ~ 0.5[23] thanks to its large intrinsic spin Hall conductivity of ~ - 2000 $\frac{\hbar}{e} s\ cm^{-1}$. A number of theoretical calculations predicted up to 40% larger spin Hall conductivities in Ta-doped β-W alloy, owing to hole doping by Ta impurities[22,24]. W, as compared to Pt, offers a better integration with standard semiconductor processing[25]. Thanks to the similarity of Ta with W, Ta/W site disorder in the A15 alloy structure is not critical for the generation of gapped Dirac-like crossings in the band structure[22], crucial for high spin Berry curvature. This also enables the use of industry-friendly sputter deposited W-Ta alloys. However, while thus promising, a systematic experimental study of spin Hall angle in W-Ta alloy to investigate the effect of Ta doping is still missing.

In this work, current-induced spin orbit torques in W$_{1-x}$Ta$_x$/Co$_{20}$Fe$_{60}$B$_{20}$/MgO systems are studied via spin-torque ferromagnetic resonance (ST-FMR) technique for Ta composition between x = 0 and x = 0.2. Both the longitudinal resistivity and spin Hall conductivity were found to vary non-monotonically with Ta concentration. A significant reduction of longitudinal resistivity to ~ 100 μΩ cm compared to a pristine β-W, paired with a large spin Hall conductivity of ~ 2800 $\frac{\hbar}{e} S\ cm^{-1}$ found at 11% Ta concentration makes this alloy one of most power efficient W-based spin Hall material for device applications.

Multilayer stacks of W$_{1-x}$Ta$_x$ (5)/ Co$_{20}$Fe$_{60}$B$_{20}$ (2)/ MgO (2) / Ta (3, thickness in nm) are sputter-deposited on naturally oxidised Si(001) substrates using a Singulus Rotaris system at room temperature. The W$_{1-x}$Ta$_x$ layers are deposited by co-sputtering of W and Ta targets and the range of Ta composition *x* is varied from 0 to 0.2. The top Ta layer oxidises completely upon exposure to air and protects the MgO layer from further oxidation. The multilayer stacks are patterned into a co-planar waveguide geometry for ST-FMR measurements using electron-beam lithography and Ar-ion milling. As shown in Fig. 1(a), the stacks are patterned into 10 µm wide and 30 µm wide strips. The Cr/Au contacts are deposited and patterned by a lift-off technique and are used as co-planar waveguides for radio frequency excitations. The measurements are performed by applying a (modulated) radio frequency signal to the sample strip (connected to the ground in the other end) and sweeping the external magnetic field with $\theta_H = 45°$ and $-135°$ while measuring the voltage across the sample strip[6]. The oscillating radio frequency current in the $W_{1-x}Ta_x$ layer generates an oscillating spin current perpendicular to the charge current direction via the spin Hall effect. The spin current enters the CoFeB layer and in turn exerts an oscillating spin torque on the CoFeB moment. When the frequency of the driving current and the magnitude of the external field satisfy the ferromagnetic resonance condition, a magnetic precession occurs. Mixing of the radio frequency current and the oscillating anisotropic magnetoresistance of the strip produces a dc voltage across the strip that is measured by a lock-in amplifier. A representative ST-FMR field sweep of 3% Ta sample is shown in Fig. 1(b). It is possible to observe FMR peaks at both positive and negative resonance fields. The peak profiles are fitted into a combination of symmetric and antisymmetric Lorentzian functions, with resonance field $H_{res}$ and half-width half maxima (HWHM) $W$ as fitting parameters. Figs. 1(c) and (d) show changes in $H_{res}$ and $W$ with the driving frequency. The change in the resonance field, $H_{res}$, with the driving frequency, $f$, is modelled using the following equation[26],

$$\mu_0 H_{res} = \frac{1}{2}\left[-\mu_0 M_{eff} + \sqrt{(\mu_0 M_{eff})^2 + 4\left(\frac{f}{\gamma}\right)^2}\right] - \mu_0 H_k, \qquad (1)$$

where $\mu_0$ is the vacuum permeability, $M_{eff}$ is the effective magnetisation, $\gamma$ is the electron gyromagnetic ratio of in CoFeB (using an electron Lande g-factor of 2.12[27]), and $H_k$ is the in-plane anisotropy field. As shown in Fig. 1(c), the model fits the data well and corresponding values of $\mu_0 M_{eff}$ and $\mu_0 H_k$ are 863 ± 30 mT and ~ 1 mT. The obtained value of $\mu_0 M_{eff}$ are similar to previous reported values for CoFeB[27,28], where the reductions from saturation magnetisation values in thin films are thought to be due to interfacial anisotropy[29]. The relationship between half-width half-maxima $W$ and the driving frequency $f$ is described by the following equation:

$$W = W_0 + \frac{2\pi\alpha}{|\gamma|}f \qquad (2)$$

where $W_0$ is the inhomogeneous broadening and $\alpha$ is the Gilbert damping constant of the CoFeB. The obtained values of $W$ follow the linear relationship well, resulting in the values of $\alpha = 0.08 \pm 0.02$ and $W_0 = 2.4 \pm 1.0$ mT. The high values of $\alpha$ in our films compared to previous literature[27–29] can result from the higher Fe composition (60% as compared to 40%) in our CoFeB films.

Current-induced spin-orbit torque measurements are performed by applying an additional bias-current to the sample[6]. The additional bias-current produces additional spin currents in the $W_{1-x}Ta_x$ layer via the spin Hall effect, which enters the neighbouring CoFeB layer. The absorbed spin current (depending on its direction with respect to the CoFeB moment direction) applies a damping-like torque to the magnetisation precession. The changes in the precession cone angle result in changes in the rectified anisotropic magnetoresistance of CoFeB, which are detected by changes in the resonance linewidth $W$. The effects of the bias-current on field sweeps are seen in Fig. 2(a), where bias-currents of up to $\pm 2$ mA both (1) shifts $H_{res}$ and (2) increases (decreases) $W$. In Fig. 2(b), the effects of a bias-current $I_{DC}$ on $W$ with different field directions of $\theta_H = 45°$ and $-135°$ are displayed. On reversing the field direction (and hence the CoFeB magnetisation direction), the sign of the damping-like torque also reverses. These changes can be interpreted as increasing/decreasing the effective Gilbert damping constant $\alpha_{eff} = \frac{|\gamma|}{2\pi f}(W - W_0)$. A linear fit through $W$ with changing bias-current density $J_C$ in the $W_{1-x}Ta_x$ layer is employed, as shown by solid lines in the same figure. The changes in the effective Gilbert damping constant $\alpha_{eff}$ with $J_C$, and subsequently the magnitude of effective damping-like torque spin Hall angle $|\theta_{DL}|$ can be obtained[26]:

$$|\theta_{DL}| = \frac{2|e|}{\hbar} \frac{\left(H_{res} + \frac{M_{eff}}{2}\right)\mu_0 M_S t_F}{|\sin\theta_H|} \left|\frac{\Delta\alpha_{eff}}{\Delta J_C}\right| \quad (3)$$

where $e$ is the electron charge, $\hbar$ is the reduced Planck constant, $M_S$ is the saturation magnetisation and $t_F$ is the CoFeB thickness. Fig. 2(c) demonstrates $|\theta_{DL}|$ with different Ta concentration, where $|\theta_{DL}|$ increases from 0.15 to 0.30 between 0 and 11% Ta concentration then decreases to 0.18 from 11% to 20%.

In order to explain this non-monotonic behaviour of $|\theta_{DL}|$ with Ta concentration, we investigate additionally two key further properties: (1) structural phases by X-ray diffraction (XRD) and (2) the longitudinal resistivity $\rho_{xx}$. Out-of-plane XRD measurements of $W_{1-x}Ta_x$ (5)/MgO (2)/Ta (3) with x = 0 – 0.2 are shown in Fig. 3(a). From the figure, three prominent peaks of $\beta$-W(210), (200) and (211) are visible, in agreement with a previous report[22,25,30]. With increasing Ta concentration, the intensities of the (200) and (211) peaks decrease. The (210) peak could be showing a small positive shift with the Ta concentration, which could be the effect of Ta-doping or a possible transformation towards the cubic α-W(110) (with a smaller $d$ of 2.23 Å). Due to the limited resolution of our data, the exact nature of shift cannot be verified. The effect of Ta-doping on the peak position partially coincides with the observed decrease in $W_{1-x}Ta_x$ film resistivity (seen in the upper plot of Fig. 3(b)) which can be interpreted as the emergence of the low-resistance alloy phases. As a reference, data from the same film stack with a pure Ta film is also shown, where the main peak is observed to be β-Ta(111).

The $W_{1-x}Ta_x$ film resistivity $\rho_{xx}$ is measured at room temperature via a four-point van der Pauw method as shown in the upper plot of Fig. 3(b). The resistivity of the 2 nm CoFeB layer is estimated by measuring samples with and without the CoFeB layer, where its longitudinal resistivity is found to be ~ 470 μΩ cm. The $W_{1-x}Ta_x$ resistivity shows an increase from 0% to 5% Ta concentration, followed by a monotonic decrease between 5% and 20%. This indicates the reduction of the high-resistance β-W phase (bulk resistivity of ~ 200 μΩ cm[23]) and emergence of the low-resistance W-Ta alloy phases with increasing Ta concentration. From the obtained values of $\theta_{DL}$ and $\rho_{xx}$, we can now compute values of transverse/spin Hall conductivity, $\sigma_{SH}$, according to $\sigma_{SH} = \theta_{DL}/\{(1 + \theta_{DL}^2)\rho_{xx}\}$ (as for high values of $\theta_{DL}$ a simple approximation of $\sigma_{SH} \approx \theta_{DL}/\rho_{xx}$ cannot be used), which is shown in the lower plot of Fig. 3(b). This shows a general increase from 0 to 11% Ta concentration with a dip at 7%, which could be due to an observed plateau/dip in the spin Hall conductivity calculations[22,24]. $\sigma_{SH}$ then decreases weakly between 11% and 20%. This trend is partially in agreement with previous *ab initio* calculations[22,24] where a 40% (60%) increase in the spin Hall conductivities was expected between 0% and 12.5% (25%) Ta concentration, where the increase was attributed mainly to effective hole doping which placed the Fermi level at the point of maximum spin Berry curvature (i.e. at the gapped band crossings). Our result suggests that such a mechanism can contribute to the trend of $\sigma_{SH}$ with Ta concentration in our alloyed system.

What is surprising in our result is that obtained values of spin Hall conductivity $\sigma_{SH}$, and damping-like spin Hall angle $\theta_{DL}$ are remarkably high, both peaking at 11% Ta concentration with values of ~ - 2800 $\frac{\hbar}{e}$ $S\ cm^{-1}$ and - 0.3, despite a low longitudinal resistivity $\rho_{xx}$ of 100 μΩ cm. With respect to reducing the power consumption, this is a significant improvement compared to pristine β-W, which has three times lower $\sigma_{SH}$ coupled with three times higher $\rho_{xx}$, as seen in Table 1. The estimated power consumption $P$ of our alloy (normalised to the value of β-W[15,28]) is the lowest of other W-based alloys such as W-Hf and W-Au, thanks to the large value of $\theta_{DL}$ with the modest value of $\rho_{xx}$. The power consumption is expected to be comparable to those in a recently published work on a Pd-Pt alloy[15], while our W-based alloy is more suitable to be incorporated to conventional device fabrication processes. The low $\rho_{xx}$ values as compared to oxygen-incorporating β-W layers[21] benefits our W-Ta alloy in terms of power consumption, despite having lower values of $\theta_{DL}$. The large spin Hall conductivity in our films as compared to other W-based alloys are thought to be due to the similarity of Ta atoms to W atoms, so that the substituted Ta atoms generate only a minimal distortion of the A15 structure and instead act as pure hole-donors. This effectively lowers the Fermi level to the area of the highest spin Berry curvature as seen in the band structure calculations[22,24].

In conclusion, we have measured current-induced spin orbit torques in $W_{1-x}Ta_x$/CoFeB/MgO/Ta systems (x between 0 and 20%) via a spin-torque ferromagnetic resonance technique. The variation in the measured effective damping-like torque and corresponding spin Hall angle as a function of Ta concentration is found to be a combined effect of changing spin Hall conductivity and longitudinal resistivity when varying the Ta concentration. The spin Hall conductivity trend suggests an increased intrinsic spin Berry curvature with increasing Ta alloying as suggested by previous theoretical work, while a small change of longitudinal resistivity suggests an intact A15 β-W structure within the studied range of Ta concentration. The large spin Hall conductivity with the low longitudinal resistivity makes this alloy system attractive for future low-power device application.

This work was funded by the Deutsche Forchungsgemeinshaft (DFG, German Research Foundation) – TRR 173 – 268565370 (projects A01 and B02) and supported by the KIST Institutional program (Project No. 2E30600). D.-S. Han acknowledges the support from the National Research Foundation of Korea (NRF) funded by the Ministry of Science and ICT (2020R1C1C1012664, 2019M3F3A1A02071509) and the National Research Council of Science & Technology (NST) (No. CAP-16-01-KIST).

[DATA AVAILABILITY]

The data that support the findings of this study are available from the corresponding author upon reasonable request.

[TABLE]

|  | $\rho_{xx}$ [$\mu\Omega\ cm$] | $\sigma_{SH}$ [$10^5 \frac{\hbar}{e} S/m$] | $\theta_{SH}$ | Normalized $P \propto \rho_{xx}/\theta_{SH}^2$ |
|---|---|---|---|---|
| β-W [28] | 300 | 1.0 | 0.3 | 1 |
| W-Hf [31] | 160 | 1.2 | 0.2 | 1.2 |
| W-Au [17] | 90 | 1.7 | 0.15 | 1.2 |
| W$_{89}$Ta$_{11}$ | 100 | 2.8 | 0.3 | 0.33 |

Table 1: Values of longitudinal resistivity $\rho_{xx}$, spin Hall conductivity $\sigma_{SH}$, maximum spin Hall angle $\theta_{SH}$ and normalized power consumption index $P \propto \rho_{xx}/\theta_{SH}^2$ of pristine β-W and W-based alloys.

[FIGURES]

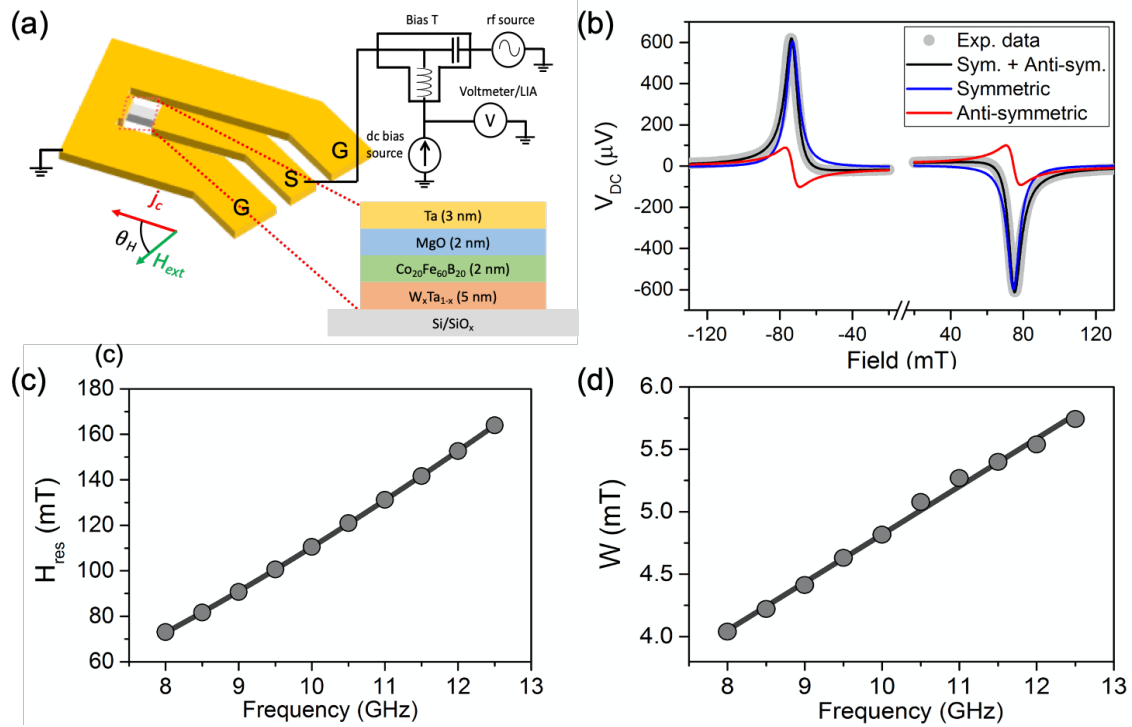

Fig. 1: (a) Schematic sample structure and ST-FMR measurement setup. (b) A representative field sweep data (grey dots) of Ta 3% sample at 8 GHz driving frequency. The fit (black line) is composed of a symmetric (blue line) and an antisymmetric (red line) Lorentzian contribution. Frequency dependence of (c) resonance field $H_{res}$ and (d) half-width half-maxima $W$.

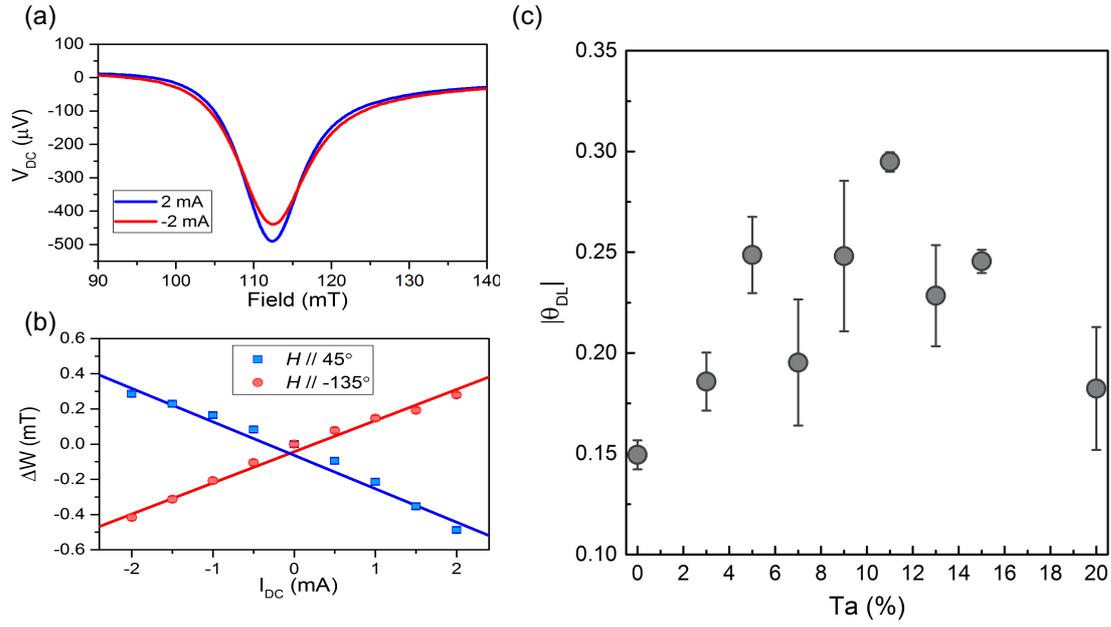

Fig. 2: (a) ST-FMR field sweep data of Ta 3% sample at 10 GHz with ±2 mA bias-current (b) Changes in half-width half-maxima $\varDelta W$ with different bias-current with field direction of $\theta_H = 45°$ and $-135°$. Solid lines are linear fit of the data. (c) Effective spin Hall angle calculated from the dependence of $W$ vs. $I_{DC}$ for different Ta concentrations.

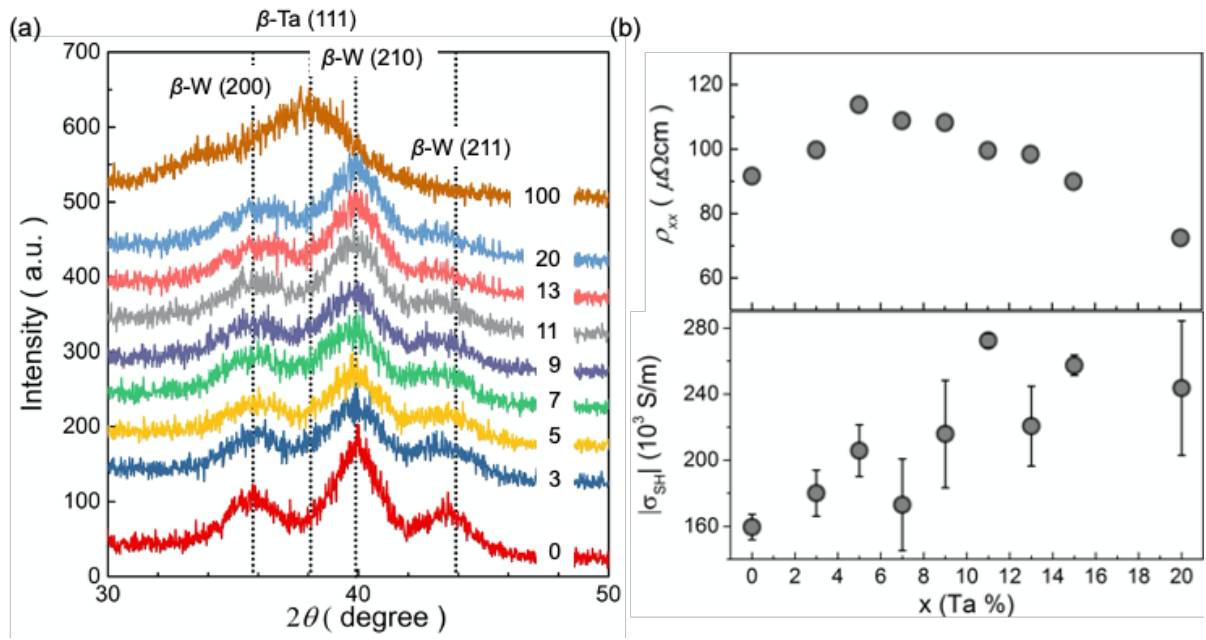

Fig. 3: (a) Out-of-plane XRD data of $W_{1-x}Ta_x$ (5) / MgO (2) / Ta (3, thickness in nm) with x = 0 – 20%. (b) Changes in longitudinal resistivity $\rho_{xx}$ and spin Hall conductivity $\sigma_{SH}$ with Ta concentration.